\documentclass[prd,aps,preprint,groupedaddress,superscriptaddress,nofootinbib]{revtex4}

\usepackage{comment}
\usepackage[utf8]{inputenc} 
\usepackage{graphicx,color,overpic,mathtools}
\usepackage{amsthm,amsmath,amssymb,hyperref,mathrsfs}
\usepackage{braket,bm,bbm,setspace}
\PassOptionsToPackage{normalem}{ulem}
\usepackage{ulem} 
\usepackage{physics}
\usepackage{float}
\usepackage[makeroom]{cancel}
\usepackage[english]{babel}
\usepackage{graphicx}
\graphicspath{ {images/} }
\addto\captionsspanish{}
\hypersetup{
    colorlinks=false,
    pdfborder={0 0 0},
}

\begin{document}

\vspace*{-2cm}
\hfill YITP-22-54
\vspace*{0.0cm}
\title{Constraints on horizonless objects after the EHT observation of Sagittarius~A*}

\author{Ra\'ul Carballo-Rubio}
\affiliation{CP3-Origins, University of Southern Denmark, Campusvej 55, DK-5230 Odense M, Denmark}
\affiliation{Florida Space Institute, University of Central Florida, 12354 Research Parkway, Partnership 1, 32826 Orlando, FL, USA}
\author{Francesco Di Filippo}
\affiliation{Center for Gravitational Physics, Yukawa Institute for Theoretical Physics, Kyoto University, Kyoto 606-8502, Japan}
\author{Stefano Liberati}
\affiliation{SISSA - International School for Advanced Studies, Via Bonomea 265, 34136 Trieste, Italy}
\affiliation{IFPU, Trieste - Institute for Fundamental Physics of the Universe, Via Beirut 2, 34014 Trieste, Italy}
\affiliation{INFN Sezione di Trieste, Via Valerio 2, 34127 Trieste, Italy}
\author{Matt Visser}
\affiliation{
School of Mathematics and Statistics, Victoria University of Wellington, PO Box 600, Wellington 6140, New Zealand
}

\begin{abstract}
The images of Sagittarius A$^*$ recently released by the Event Horizon Telescope collaboration have been accompanied [Ap.J.Lett.\,{\bf 930\,\#2}\,(2022)\,L17]
by an analysis of the constraints on the possible absence of a trapping horizon, i.e.~on the possibility that the object at the center of our galaxy is an ultra-compact object with a surface re-emitting incident radiation. Using the observed image size and the broadband spectrum of Sgr A$^*$, it is claimed that the radius of any surface, in which incident radiation is re-emitted thermally, is strongly bounded from above by these latest observations. Herein, we discuss how the reported constraint relies on the extremely strong assumption of perfect balance in the energy exchange between the accretion disk and the central object, and show that this is violated whenever the surface is endowed with any  non-zero absorption coefficient. We derive the upper-bound constraints that can be cast on the radius and dimensionless absorption coefficient of the surface. We show that the conclusions of the analysis presented by the EHT collaboration hold only for unnaturally small values of the absorption coefficient (i.e. much lower than $10^{-14}$), and thus have to be significantly revised in scenarios with physical significance. 
\end{abstract}

\maketitle
\def\O{{\mathcal{O}}}

\section{Introduction}

Black holes have played a central role in our understanding and testing of gravity since the early years of General Relativity (GR). These elegant exact solutions have exhibited a particularly rich wealth of theoretical and phenomenological issues that have propelled intense research for over a century. Nonetheless, their simplicity is also associated with ``extreme" structures such as singularities, event horizons, and Cauchy horizons, which have stimulated the investigation of possible alternative endpoints for gravitational collapse~\cite{bh-micro-survey}. Indeed, the current literature is characterized by a whole range of mimickers of GR black holes, such as regular (singularity free) trapped regions, bouncing solutions, wormholes or even ultra-compact horizonless objects (see~\cite{Carballo-Rubio:2019fnb,Carballo-Rubio:2019nel} for a classification of possible scenarios).

In the absence of direct observations, such black hole mimicking  compact objects have for many years remained  just mere speculation, amenable only to theoretical investigation. However, the recent direct observations of gravitational waves and VLBI images have reignited the field, opening the possibility of testing a whole new branch of alternative theories of gravity in what can be considered a new kind of quantum gravity phenomenology, given that some of the above-mentioned compact objects are sometimes directly linked to singularity regularization induced by quantum gravity scenarios. (See e.g.~\cite{Cardoso:2017,Carballo2018,Cardoso:2019rvt} for a comprehensive discussion concerning the phenomenology and constraints on a broad class of black hole mimickers.) As part of the release of their Sagittarius (Sgr) A$^*$ results~\cite{EventHorizonTelescopeI,EventHorizonTelescopeII,EventHorizonTelescopeIII,EventHorizonTelescopeIV,EventHorizonTelescopeV,EventHorizonTelescopeVI}, one of the papers recently published by the Event Horizon Telescope (EHT) collaboration~\cite{EventHorizonTelescopeVI} focuses on the study of the constraints that can be placed on a specific class of black hole mimickers ---  implicitly providing additional evidence of the considerable interest in these alternatives within the communities involved --- i.e. ultra-compact horizonless objects.

\section{Constraining ultra-compact horizonless objects}

Ultra-compact horizonless objects are a class of black hole mimickers characterized by a physical surface hovering just above the would-be trapping horizon, and well within the photon sphere characterizing the electromagnetic shadow of a black hole candidate. There are many proposals for such an object, and we provide here a sample of references without the aim of being exhaustive~\cite{Chapline:2000en,Amati:2000ne,Mazur:2004fk,gravastars, anisotropic,Mathur:2005zp,Vachaspati:2006ki,Amati:2006fr,small-dark-heavy,Barcelo:2010vc,Brustein:2016msz,Carballo-Rubio:2017tlh,Arrechea:2021pvg}. (See the more complete discussions in e.g.~\cite{Cardoso:2017,Carballo2018}). To be even more precise, we shall here assume the following assumptions, satisfied by all models in this class of black hole mimickers:\\ (i) the geometry is Schwarzschild above a given radius $R_*$ that is defined to be the radius of the object, (ii) the geometry for $r \leq R_*$ is not Schwarzschild, and (iii) there are no event or trapping horizons,\footnote{Note that event horizons are not observable by definition, but quasi-local notions such as trapping horizons are \cite{observability}. We advocate here for the use of quasi-local definitions of horizons and thus of black holes, as event horizons are not necessary to devise black hole physics~\cite{Ashtekar:2004cn,Booth:2005qc,production-and-decay}.} but a surface endowed with physical properties, including the possibility to absorb but not completely re-emit impinging radiation.

Of course, the presence of a physical surface so close to the would-be horizon of a compact object should not be taken lightly. It is not the scope of this brief paper to discuss all possible models associated to such geometries, which of course in any case would have to envisage some form of new physics. Regarding this aspect, we shall here be agnostic and simply discuss casting constraints on the presence (or absence) of such a surface, which is a standard approach in the literature~\cite{Narayan:1997xv,Garcia:2000ew,Abramowicz:2002vt,Broderick:2005,Broderick:2009ph,Broderick:2015tda,Lu:2017vdx,Cardoso:2017,Carballo-Rubio:2018vin,Carballo2018,Cardoso:2019rvt,EventHorizonTelescopeVI}.

It is then useful to introduce different phenomenological coefficients that parametrize the physical processes taking place during the interaction between photons and the surface~\cite{Carballo2018}. Photons can be elastically reflected, absorbed and re-emitted as thermal radiation, or absorbed by the bulk degrees of freedom of the object. We can define dimensionless coefficients associated with these three channels, defined as the ratio of the energy that goes into each channel with respect to the total incident energy, thus having $\Gamma$ (elastic channel), $\tilde{\Gamma}$ (inelastic or thermal channel), and $\kappa$ (absorption). We will assume that the bulk of the central object is optically thick, which translates into the sum rule $\Gamma+\tilde{\Gamma}+\kappa=1$.

To date, the most stringent constraints on the viability of such quasi-black hole (horizon\-less) solutions come from information about the average amount of infalling matter per unit time impinging on the closest supermassive black hole candidate, i.e.~our own galactic supermassive black hole Sgr A$^*$. Both upper bounds~\cite{Broderick:2005,Broderick:2009ph,EventHorizonTelescopeVI} and lower bounds~\cite{Carballo-Rubio:2018vin} on the size of a hypothetical surface have been reported, which leaves some parameter space for viable horizonless configurations~\cite{Carballo2018}. Lower and upper bounds stem from different arguments and are thus conceptually independent from each other (in particular, lower bounds do not depend on the luminosity of the surface, and are therefore not changed by considerations regarding absorption). Upper-bound constraints rely on the existence of inelastic processes (thermal emission), and are therefore meaningful if and only if $\tilde{\Gamma}\neq 0$. In~\cite{EventHorizonTelescopeVI}, the Event Horizon Telescope collaboration has reported upper-bound constraints (\emph{Sec. 4.1 - Thermalizing Surface}), focusing on situations in which $\Gamma=0$ and $\tilde{\Gamma}=1$, thus ignoring absorption completely.\footnote{For completeness, let us mention that~\cite{EventHorizonTelescopeVI} contains a discussion (\emph{Sec. 4.2 - Reflecting Surface}) of simulated images for specific models in which the opposite situation is realized, that is, only elastic reflection plays a role. The discussion of these images explicitly includes absorption, thus covering situations in which $\tilde{\Gamma}=0$ and $0\leq \Gamma\leq 1$.} In this short paper we focus on critically assessing these upper-bound constraints. We will thus focus on the inelastic (thermal) channel, considering for simplicity that $\Gamma=0$, but considering explicitly the role of absorption through the relation $\tilde{\Gamma}=1-\kappa$.
 
We can summarize the main argument leading to upper-bound constraints for the inelastic (thermal) channel~\cite{Broderick:2005,Broderick:2009ph,Broderick:2015tda} (see also~\cite{Abramowicz:2002vt,Cardoso:2017,Carballo-Rubio:2018vin,Cardoso:2019rvt}) as follows: Let us consider the system composed of the black hole candidate and its accretion disk. In such a system energy is continuously exchanged between its two components: the accretion rate $\dot{M}$ quantifies the flux of energy accreting from the disk onto the ultra-compact object; while, in the presence of a surface, one has to expect a flux of energy $\dot{E}$ re-emitted from the central object and getting back to the accretion disk. A constraint on the presence of a surface can then be cast by an observation providing an upper bound on the ratio $\dot{E}/\dot{M}$. 

Now, in order to do this directly, one would need to disentangle the two fluxes of energy and measure them independently. However, this is not yet observationally possible. 
Therefore, current constraints have to be based on a set of additional assumptions that help to further characterize the properties of the outgoing energy flux.
In particular, in line with the previous literature on the subject, the analysis of Ref.~\cite{EventHorizonTelescopeVI}, considers the following assumptions:
\vspace{-7pt}
\begin{itemize}
\itemsep-5pt
    \item[\textit{i.}] The matter in the compact object at the center of Sgr A* satisfies energy conservation; 
    \item[\textit{ii.}] It couples to, and radiates in, all electromagnetic modes;
    \item[\textit{iii.}] The system (compact object plus accretion disk and their reciprocal fluxes) obeys the laws of thermodynamics and, in particular, approaches statistical equilibrium in a steady 
    state.
\end{itemize}
\vspace{-7pt}
\enlargethispage{10pt}
In this short note, we will comment on the validity of this last assumption. There are two physical mechanisms that can lead to significant violations of this assumption: strong lensing and non-zero absorption. The first of these mechanisms has been (partially) taken into account in Sec. 4.1 of~\cite{EventHorizonTelescopeVI}; however, the effect of non-zero absorption coefficient $\kappa$ has been overlooked when analyzing thermal emission, as we have mentioned above. In particular, in this brief paper we shall stress, providing further details regarding the discussion in~\cite{Carballo2018}, that even a very small absorption coefficient can vitiate this assumption, thus significantly weakening the subsequent constraint.\footnote{Additionally, it is reasonable that only a small fraction of the energy radiates by the surface will be carried away by photons, so that other particles are also produced. This will be discussed elsewhere~\cite{Carballo-Rubio:2022xyz}.} 

\section{Surface luminosity as a function of absorption}

We are considering, for simplicity, a spherically symmetric object of mass $M$ and radius $R_*$. In this section, we calculate the (thermal) luminosity associated with such an object, under the assumptions introduced in the previous section. Define the quantity
\begin{equation}
    \mu:=1-\frac{2M}{R_*}\,,
\end{equation}
which is related to the so-called compactness $\chi=2M/R_*$. 
As pointed out in Ref.~\cite{EventHorizonTelescopeVI}, the minimum surface luminosity expected at infinity $L_\infty$ can be estimated as\footnote{Throughout this paper we use units in which $G=c=1$.}
\begin{equation}
    L_\infty>\eta\; \dot{M}\,,
\end{equation}
where $\dot{M}$ is the accretion rate, and
\begin{equation}
\eta:=\frac{\dot{E}}{\dot{M}}    
\end{equation}
is the fraction of the accretion mass that is released to spatial infinity. 

An upper bound on the observed luminosity can then be translated into a constraint on the $\eta$ parameter. 
From the right panel of Fig.~14 of~\cite{EventHorizonTelescopeVI}, we can infer that the observed luminosity is roughly two orders of magnitude smaller than the predicted luminosity for a conservative value of the accretion flow. Therefore, we will consider the conservative constraint
\begin{equation}
    \eta\lesssim 10^{-2}\,.
\end{equation}
In in order to translate such a constraint into one for the object's $\mu$, one needs to find a relation between $\mu$ and $\eta$. 
In~\cite{Broderick:2005}, and Sec. 4.1 of~\cite{EventHorizonTelescopeVI}, assuming that all the kinetic energy of infalling matter is converted to outgoing radiation leads to the naive result
\begin{equation}\label{eq:etanaive}
\eta = 1-\left( 1-\frac{2M}{R_*} \right)^{1/2}=1-\sqrt{\mu}\,.
\end{equation}
However, as we will see, the real expression for $\eta$ is very different from this naive estimate. In fact, Eq.~\eqref{eq:etanaive} does not take into account two very important physical mechanisms --- strong lensing and non-zero absorption.

\begin{figure}[!htbp]
\vbox{\includegraphics[scale=.5]{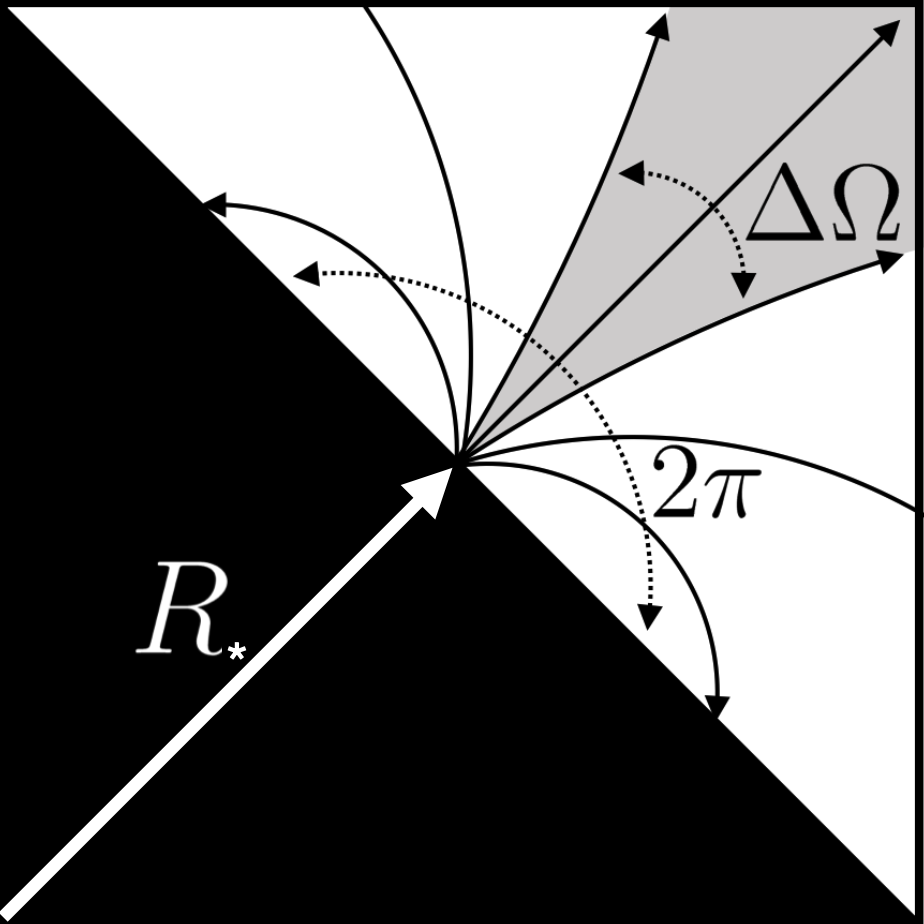}}
\caption{When radiation hits the surface of the compact object at best a fraction $\Delta\Omega/2\pi$ escapes. The onset of steady-state accretion can be significantly delayed by having part of the accretion flow bounce back and forth multiple times between the surface $R_*$ of the compact object and the photon sphere at $r=3M$. This behavior is further changed once including absorption. 
\label{F-beam}
}
\end{figure}

If $R_* < 3M$ (that is, if the surface of the compact object lies below the photon sphere), then
matter that is emitted at the surface can escape the gravitational field and reach spatial infinity only  if it is emitted within a narrow solid angle \cite{Abramowicz:2002vt,Lu:2017vdx,Cardoso:2017,Carballo2018}
(see figure \ref{F-beam}): 

\begin{equation}\label{eq:esc_ang}
    \frac{\Delta\Omega}{2\pi}
    =  1 + \left(1-\frac{3M}{R_*}\right) \sqrt{1+\frac{6M}{R_*}}
    =
    1 - (1-3\mu)\sqrt{1-\frac{3\mu}{4}} 
    = \frac{27}{8}\, \mu +\O(\mu^2)\,.
\end{equation}
%
Because of this beaming effect,  if $\mu$ is small enough the radiation will bounce back onto the surface several times, causing a significant delay in the onset of the steady state regime \cite{Cardoso:2017,Carballo2018}. 

Furthermore, it is unrealistic to assume that all the energy absorbed by the surface is immediately re-emitted back. It is reasonable to expect that at least some absorption of energy will be taking place, due to the internal degrees of freedom of the central object. In the following, we will denote with $\kappa$ the absorption coefficient, which measures the fraction of energy that is (semi)permanently lost inside the compact object.
For a black hole we have $\kappa=1$. The absence of a trapping horizon does not necessarily imply that $\kappa=0$ so that all the electromagnetic energy that hits the surface are necessarily re-emitted. 
This will turn out to be particularly important, given the large number of times the radiation bounces off the surface of the object. As we will quantify in detail below, this ultimately implies that even a very small absorption coefficient can have a highly non-negligible effect.

To show this, let us consider discrete intervals of time, with size given by the characteristic timescale 
\begin{equation}
    \tau=\mathcal{O}(10)M\sim10^2\mbox{ s}
\end{equation} 
between two consecutive bounces on the surface, which is of the same order as the time necessary for the matter to reach the surface of the compact object from the accretion disk. During each of
these time intervals, the mass that the accretion disk is ejecting into the compact object is given by $\dot{M}\,\tau$. We denote with $E_n$ the amount of energy that reaches the accretion disk after $n$ time intervals. It will also prove useful to define the energy $\epsilon_n$ that reaches the accretion disk after bouncing off the compact object $n$ times.

\begin{figure}
\begin{center}
\vbox{\includegraphics[scale=.6]{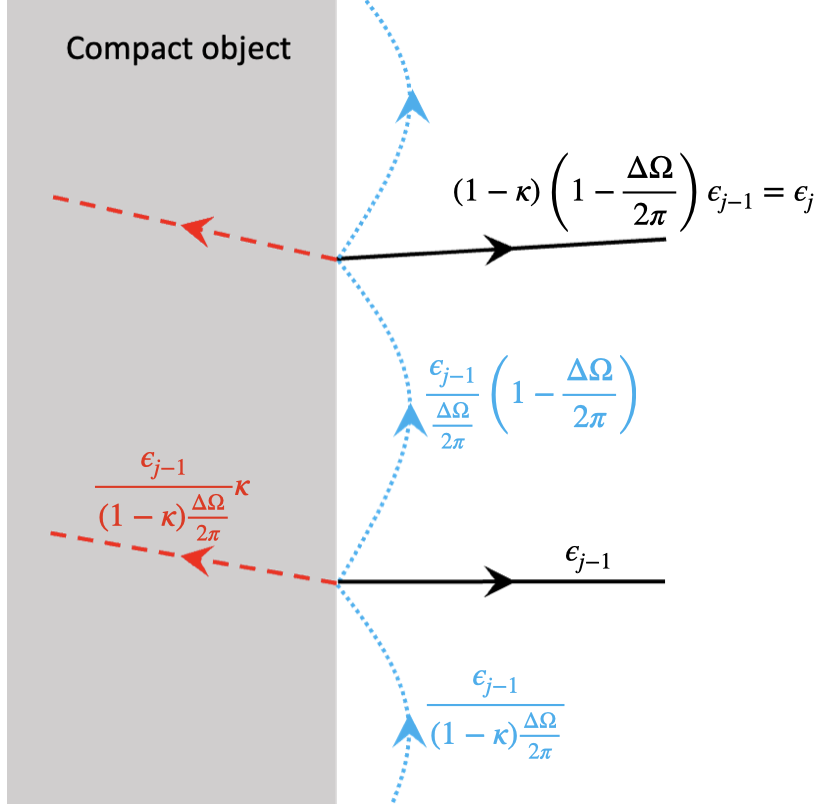}}
\caption{Schematic proof of Eq.~\eqref{eq:5}. When some radiation hits the surface of the compact object (blue dotted lines) a fraction $(1-\kappa)\Delta\Omega/2\pi$ escapes (black lines), a fraction $\kappa $ is absorbed (red dashed lines) and the rest falls back onto the surface of the compact object.
\label{F-schematic}}
\end{center}
\end{figure}

The energy that escapes at infinity after $n$ time intervals is given by
\begin{equation}
 E_n=\sum_{j=1}^n\epsilon_j
\end{equation} 
 We note that $\epsilon_1=\left( 1-\kappa  \right)\frac{\Delta\Omega}{2\pi}\dot{M}\tau$ and (see Fig.~\ref{F-schematic}):
\begin{equation}\label{eq:5}
\epsilon_j=\left(1-\kappa\right)\left(1-\frac{\Delta\Omega}{2\pi}\right)\epsilon_{j-1}.
\end{equation}
So,
\begin{equation}
\epsilon_j=\left(1-\kappa\right)^j\left(1-\frac{\Delta\Omega}{2\pi}\right)^{j-1}\frac{\Delta\Omega}{2\pi}\dot{M}\tau.
\end{equation}
Therefore, we have a geometric series which can easily be summed:
\begin{equation}
\begin{array}{rl}

\displaystyle\frac{E_n}{\dot{M}\tau}=&\displaystyle \frac{\Delta\Omega}{2\pi} (1-\kappa)\sum_{j=0}^{n-1}\left\{  (1-\kappa)^j\left(1-\frac{\Delta\Omega}{2\pi}\right)^j	\right\}\\
\\
=&\displaystyle\frac{\Delta\Omega}{2\pi}\frac{(1-\kappa)}{\kappa+\frac{\Delta\Omega}{2\pi}(1-\kappa)}\left\{ 1-(1-\kappa)^n\left(1-\frac{\Delta\Omega}{2\pi}\right)^n	\right\}.

\end{array}
\end{equation}
Going back to the continuum limit, we have
\begin{equation}\label{eq:etatot}
    \eta(t)=\frac{\dot{E}}{\dot{M}}=\frac{\Delta\Omega}{2\pi}\frac{(1-\kappa)}{\kappa+\frac{\Delta\Omega}{2\pi}(1-\kappa)}\left\{ 1-(1-\kappa)^{t/\tau}\left(1-\frac{\Delta\Omega}{2\pi}\right)^{t/\tau}	\right\}.
\end{equation}
Here, $t$ denotes the timescale over which $Sgr\,A^*$ has been accreting. One possible estimate for this timescale~\cite{Broderick:2005,Lu:2017vdx,Carballo-Rubio:2018vin} is given by the Eddington timescale\footnote{We will see that the precise value of the timescale is not particularly relevant for generic non-vanishing values of $\kappa$.}, $t\sim T\sim3.8\cross10^8\mbox{ yr}$.

Eq.~\eqref{eq:etatot} is the main equation in the paper. It supersedes the naive estimation in Eq.~\eqref{eq:etanaive}, as it adequately describes the effect of gravitational lensing as well as absorption. We will devote the rest of the paper to extracting the physics encoded in this equation.

\begin{figure}[!htbp]
\vbox{\includegraphics[scale=.55]{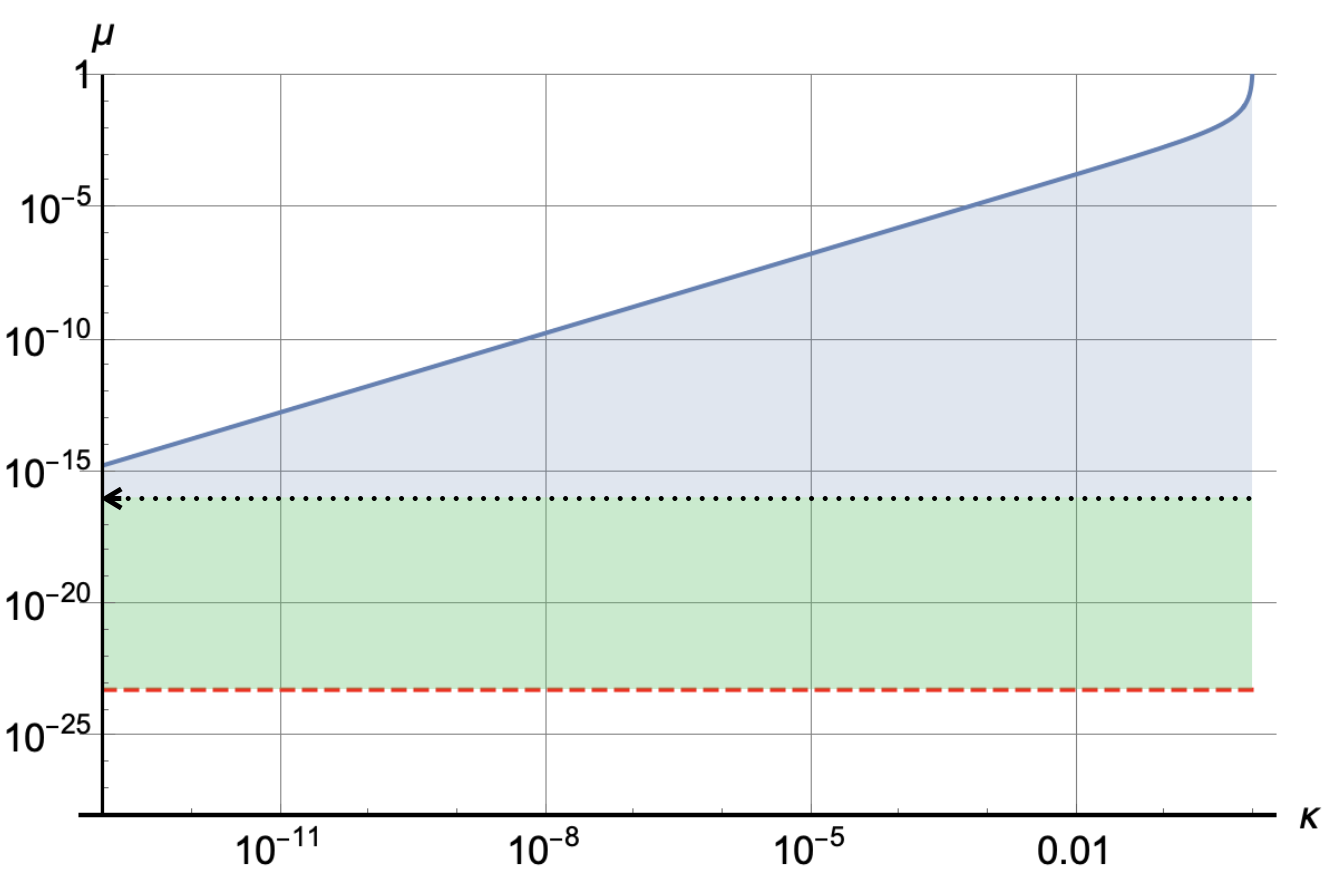}}
\caption{Plot of the allowed values of $\mu$ as a function of $\kappa$ in Log-Log scale. The allowed values are in the shaded area, between the solid blue line (upper-bound constraint taking into account absorption) and the dashed red line (lower-bound constraint from causality arguments~\cite{Carballo-Rubio:2018vin}, that we have not discussed here but is included for completeness). The dotted black line corresponds to the naive extrapolation of the upper-bound constraint that ignores absorption. Even very small values of the absorption coefficient (e.g. $10^{-5}$) significantly weaken the upper-bound constraint by several orders of magnitude. Viable values of the compactness of horizonless objects span several orders of magnitude (up to around 20, depending on the value of absorption).
\label{Fig:Constraint}
}
\end{figure}

\begin{figure}[!htbp]
\vbox{\includegraphics[scale=.6]{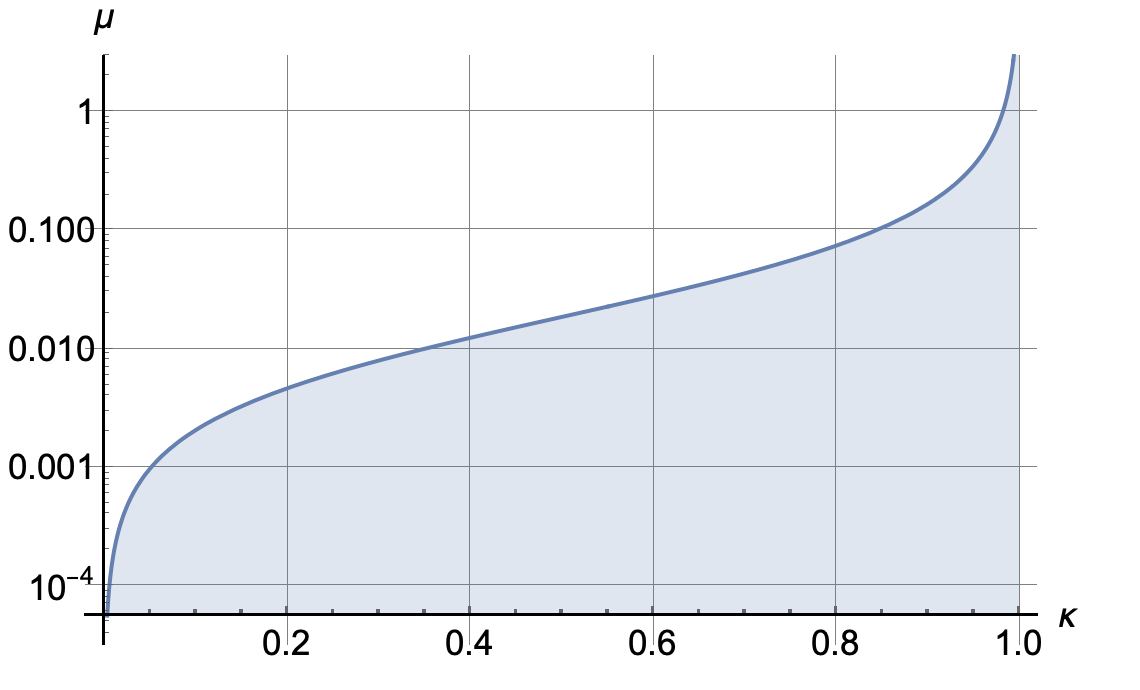}}
\caption{Plot showing the same quantities as in Fig.~\ref{Fig:Constraint}, but in Log-Linear scale to better display higher values of the absorption coefficient. This figure illustrates that no meaningful upper-bound constraints can be placed, using current observations, on horizonless objects that are sufficiently close to displaying perfect absorption. The dashed-red and dotted-black lines in Fig.~\ref{Fig:Constraint} are not visible in Log-Linear scale.
\label{Fig:Constraint2}
}
\end{figure}

We can identify two different regimes, depending on the value of the absorption coefficient:
\begin{itemize}
    \item $\kappa\ll \tau/T\ll 1$ (which includes the particular case $\kappa=0$ considered in Sec. 4.1 of~\cite{EventHorizonTelescopeVI}): using the relation
    \begin{equation}
    \left(1-\kappa\right)^{T/\tau}=e^{T\ln\left(1-\kappa\right)/\tau}\simeq e^{-\kappa T/\tau}\simeq 1,
\end{equation}
    it follows that the term proportional to $    \left(1-\kappa\right)^{T/\tau}$ in Eq.~\eqref{eq:etatot} approximately equals unity, so that we can write
    \begin{equation}\label{etak0}
    \eta(T)\simeq 1-\left(1-\frac{\Delta\Omega}{2\pi}\right)^{T/\tau}\,.
\end{equation}
The constraint $\eta\lesssim10^{-2}$ translates into the upper bound
\begin{equation}\label{eq:nc1}
\mu\lesssim 10^{-16}\,.    
\end{equation}
There are two very important points to note about this constraint. First of all, this is actually a tighter constraint than the one ($\mu<10^{-14} $) reported in Sec.~4.1.4 of~\cite{EventHorizonTelescopeVI}, as in that paper the constraint is obtained by imposing that the steady state $\eta\sim 1$ is not reached,  instead of $\eta\lesssim10^{-2}$ that we are imposing here. More importantly, this constrain relies on the assumption that $\kappa$ either vanishes identically or is extremely small, which is arguably unrealistic given the ubiquitous presence of absorption in astrophysical processes. If is, therefore, much more reasonable to consider the regime described next.

    \item $\tau/T\ll \kappa< 1$: using the relation
    \begin{equation}
    \left(1-\kappa\right)^{T/\tau}=e^{T\ln\left(1-\kappa\right)/\tau}< e^{-\kappa T/\tau}\ll 1\,,
\end{equation}
where we have used the inequality
\begin{equation}
    \ln{(1-x)}<-x\qquad\forall \,x \in \,\,(0,1)\,.
\end{equation}
It follows that the term proportional to $  \left(1-\kappa\right)^{T/\tau}$ in Eq.~\eqref{eq:etatot} is negligible, so that we can write
    \begin{equation}\label{eq:constrkneq0}
\eta=\frac{\Delta\Omega}{2\pi}\;\frac{1-\kappa}{\kappa+\frac{ \Delta\Omega}{2\pi}(1-\kappa)}.
\end{equation}
This result is very different from Eq.~\eqref{etak0}. First of all, note that the time dependence has disappeared. Also, $\dot{E}\neq\dot{M}$, which is reasonable to expect as part of the energy deposited onto the surface is absorbed. We can see that Eq.~\eqref{eq:constrkneq0}, once compared with observations, cast a much weaker constraint (see Fig.~\ref{Fig:Constraint}). For example, an absorption coefficient as small as $\kappa=10^{-5}$ weakens the the upper-bound constraint to 
\begin{equation}
\mu\lesssim10^{-7},\qquad \left(\kappa=10^{-5}\right).
\end{equation}
This is 9 order of magnitudes weaker than the naive constraint in Eq.~\eqref{eq:nc1}. As another example, a value of $1-\kappa=10^{-2}$ leads to the constraint
\begin{equation}
\mu\lesssim 1,\qquad \left(1-\kappa=10^{-2}\right).
\end{equation}
\enlargethispage{15pt}
This implies that no meaningful upper-bound constraints can be placed for objects that are sufficiently close to displaying perfect absorption, as Fig.~\ref{Fig:Constraint2} illustrates more clearly.

\end{itemize}

\section{Discussion and conclusions}

In this short paper, we have discussed constraints that can be cast on horizonless alternatives to standard black holes via electromagnetic observations. Following the discussion of thermal emission in~\cite{EventHorizonTelescopeVI}, we have focused on the spherically symmetric case for simplicity. It is known that, in the rotating case, the escape angle in Eq.~\eqref{eq:esc_ang} would depend on the latitude. However, as shown in \cite{Zulianello:2020cmx}, for an object like Sgr A* for which the angular momentum is below $a\simeq0.9$, the emission is very nearly isotropic and we can simply substitute the latitude dependent escape angle with its average value. 

The main message of this paper is that the upper-bound constraint for thermal emission reported by the EHT collaboration in~\cite{EventHorizonTelescopeVI} only partially accounts for the strong lensing effect and, crucially, it does not consider the effect of an absorption coefficient. We have shown that this is far from being a justified working assumption, as even a very small absorption coefficient renders the assumptions in Sec.~4.1 of~\cite{EventHorizonTelescopeVI} invalid. The revised discussion presented here, that accounts for absorption, results into an upper-bound constraint that is generally weaker by several orders of magnitude, even for absorption coefficients as small as $10^{-5}$. This revised analysis leaves large and physically-motivated regions of parameter space unconstrained.

Let us stress again that this result does not require exotic new physics. The bending of light is obtained within a standard general relativistic framework and its extreme behaviour is only to be expected in ultra-compact objects. Similarly, the expectation of a non-zero, albeit small, absorption coefficient has to be expected on physical grounds. In fact, reversing the logic of our statement, an absorption coefficient that vanishes exactly would probably have to entail a quite odd behavior (and thus some form of new physics) of the surface.

In conclusion, we think that this should serve as a cautionary note concerning the fact that the physics of ultra-compact horizonless objects is still relatively unknown, and that analyses that ignore some of its subtleties may provide us with an ill-posed over-confidence in having strongly constrained such objects with current observations. We believe that, as illustrated in this paper, more refined analyses can provide us with additional insights regarding these constraints.

\acknowledgments 
We are indebted to Luciano Rezzolla and Sebastian H. V\"olkel for helpful comments on the first version of the manuscript. RCR acknowledges financial support through a research grant (29405) from VILLUM fonden. FDF acknowledges financial support by Japan Society for the Promotion of Science Grants-in-Aid for international research fellow No. 21P21318. 
SL acknowledges funding from the Italian Ministry of Education and  Scientific Research (MIUR)  under the grant  PRIN MIUR 2017-MB8AEZ. 
MV was supported by the Marsden Fund, via a grant administered by the Royal Society of New Zealand. 

\bibliography{refs}

\end{document}